\documentclass[12pt,titlepage]{article}
\usepackage{graphicx}
\usepackage{amsmath}
\usepackage{amssymb}
\usepackage{latexsym}
\usepackage{xcolor}
\usepackage{multirow}
\usepackage[round,sort&compress,comma,numbers]{natbib}
\usepackage{fancyhdr}
\usepackage[small,bf]{caption}

\makeatletter \renewcommand{\@biblabel}[1]{#1.} \makeatother

\advance\voffset by -1cm \advance\hoffset by -1.2cm
\iffloatpage{\fancyfoot[C]{}}{\fancyfoot[C]{\thepage}}

\newcommand{\bii}{\begin{itemize}}
\newcommand{\eii}{\end{itemize}}
\newcommand{\bie}{\begin{enumerate}}
\newcommand{\eie}{\end{enumerate}}
\newcommand{\beq}{\begin{equation}}
\newcommand{\ee}{\end{equation}}
\newcommand{\eeq}{\end{equation}}
\newcommand{\beqa}{\begin{eqnarray}}
\newcommand{\eea}{\end{eqnarray}}

\newcommand{\no}{\noindent}

\newcommand{\al}{\alpha}

\newcommand{\bc}{\begin{center}}
\newcommand{\ec}{\end{center}}

\newcommand{\fref}[1]{Figure\ \ref{fig:#1}}
\newcommand{\tref}[1]{Table\ \ref{tab:#1}}

\newcommand{\eref}[1]{eqn.\ (\ref{eqn:#1})}

\newcommand{\id}{\mathrm{d}}
\newcommand{\Dt}[1]{\frac{\id #1}{\id t}}

\newcommand{\vivo}{{\it in vivo\ }}

\newcommand{\mat}{\begin{array}{cc}}
\newcommand{\emat}{\end{array}}

\newcommand{\ra}{\rightarrow}

\newcommand{\LD}{\LD_{50}}

\newcommand{\arw}[1]{\stackrel{#1}{\rightarrow}}
\newcommand{\arws}[2]{\overset{#1}{\underset{#2}{\rightleftharpoons}}}

\graphicspath{{/home/fly10/vitaly/research/pictures/}{figs/}}

\renewcommand{\cite}{\citep}

\newcommand{\eps}{\epsilon} %
\newcommand{\freq}{\mathcal{T}}

\begin{document}

\title{Killing of targets by effector CD8$^+$T cells in the mouse spleen
  follows the law of mass action}

\author{{Vitaly V. Ganusov$^{1,3}$, Daniel L. Barber$^2$, and Rob J.
    De Boer$^3$}
  \\
  {\small $^1$Los Alamos National Laboratory, Los Alamos, NM 87505, USA}\\
  {\small $^2$National Institutes of Health, 9000 Rockville Pike,
    Bethesda, MA 20892 USA}\\
  {\small $^3$Theoretical Biology, Utrecht University, 3584 CH
    Utrecht,  The Netherlands}\\
  {\small Email: vitaly.ganusov@gmail.com}
                                %
}

\maketitle

\begin{abstract}
  It has been difficult to measure efficacy of T cell-based vaccines
  and to correlate efficacy of CD8$^+$T cell responses with protection
  against viral infections. In part, this difficulty is due to our
  poor understanding of the \vivo efficacy of CD8$^+$T cells.  Using a
  recently developed experimental method of \vivo cytotoxicity we
  investigated quantitative aspects of killing of peptide-pulsed
  targets by effector and memory CD8$^+$T cells, specific to three
  epitopes of lymphocytic choriomeningitis virus (LCMV), in the mouse
  spleen.  By analyzing data on killing of targets with varying number
  of epitope-specific effector and memory CD8$^+$T cells, we find that
  killing of targets by effectors follows the law of mass-action, that
  is the death rate of peptide-pulsed targets is proportional to the
  frequency of CTLs in the spleen.  In contrast, killing of targets by
  memory CD8$^+$T cells does not follow the mass action law because the
  death rate of targets saturates at high frequencies of memory CD8$^+$T
  cells.  For both effector and memory cells, we also find no support
  for a killing term that includes the decrease of the death rate of
  targets with increasing target cell density.  Importantly, we find
  that at low CD8$^+$T cell frequencies, effector and memory CD8$^+$T cells
  on the per capita basis are equally efficient at killing
  peptide-pulsed targets.  Our framework provides the guideline for
  the calculation of the level of memory CD8$^+$T cells required to
  provide sterilizing protection against viral infection.  Our results
  thus form a basis for quantitative understanding of the process of
  killing of virus-infected cells by T cell responses in tissues and
  can be used to correlate the phenotype of vaccine-induced memory CD8
  T cells with their killing efficacy \vivo.

\vspace{0.2cm}
 
 Short running title: Killing by CD8$^+$T cells in tissues
 
 Abbreviations: LCMV, lymphocytic choriomeningitis virus, CTLs,
 cytotoxic T lymphocytes, CIs, confidence intervals, RSS, residual sum
 of squares

\end{abstract}

\section{Introduction}

Vaccination is often considered as one of the greatest medical
achievements of the last century but due our limited understanding of
the correlates of protection, most vaccines have been developed by a
trial and error approach and we have failed to deliver vaccines for
important diseases like AIDS or malaria. It is generally believed that
most of the currently used vaccines provide protection by inducing
high titers of pathogen-neutralizing antibodies \cite{Pantaleo.nm04}.
The efficacy of an antibody-inducing vaccine is generally proportional
to the titer of neutralizing antibodies after vaccination
\cite{Plotkin.cid08}.  Several vaccines that are currently being
developed for chronic infections such as HIV and malaria, are aimed to
stimulate T cell responses. It is unclear, however, what parameters of
the T cell memory that is induced by vaccination, best correlate with
protection \cite{Pantaleo.nm04}. It has been suggested that
polyfunctional memory CD4 T cells may be superior in providing
protection following infection with Leishmania \cite{Darrah.nm07} and
polyfunctional memory CD8$^+$T cells are protective against SIV infection
\cite{Liu.n09}, but for some important human infections, such as HIV
infection, evidence is still lacking \cite{Rehr.jv08,Streeck.pm08}.

In part, our limited understanding of how memory T cells provide
protection comes from the fact that most effector functions of
effector and memory T cells are measured in vitro (after short- or
long-term restimulation), and there is very little quantitative
details of how T cells control pathogen growth in tissues (e.g.,
\cite{Mempel.i06b}). Recently, a new experimental technique to measure
cytotoxic efficacy of CD8$^+$T cells \vivo has been introduced
\cite{Aichele.i97,Oehen.ji98,Barchet.eji00,Mueller.jem02,Coles.ji02}.
In this assay, peptide-pulsed and unpulsed target cells are
transferred into mice harboring peptide-specific effector or memory
CD8$^+$T cells, and elimination of pulsed targets is used as indication
of Ag-specific killing \vivo
\cite{Byers.ji03,Barber.ji03,Curtsinger.jem03,Hermans.jim04,Ingulli.mmb07}.
We use the data from recently published experiments \cite{Barber.ji03}
and use a recently developed mathematical model
\cite{Regoes.pnas07,Yates.po07,Ganusov.jv08} to investigate
quantitative details how of effector and memory CD8$^+$T cells, specific
for three epitopes of lymphocytic choriomeningitis virus (LCMV), kill
peptide-pulsed targets in the mouse spleen.

Unexpectedly, our results suggest that killing of targets by effector
CD8$^+$T cells (present at the peak of the immune response) follows the
law of mass action: the rate of killing is simply proportional to the
density of targets and the frequency of effector CD8$^+$T cells in the
spleen. Such a linear dependence of the death rate of targets on the
frequency of effectors was observed over 100 fold range of effector
frequencies.  In contrast, killing of targets by LCMV-specific memory
CD8$^+$T cells does not follow the law of mass action as the death rate
of peptide-pulsed targets saturates at high frequencies of memory CD8
T cells. This saturation suggests that there might be an upper bound
level of efficacy of the total memory T cell response, and this may
potentially limit the efficacy of T-cell based vaccines.
Interestingly, we found that at low CD8$^+$T cell frequencies, effector
and memory CD8$^+$T cells are equally efficient at clearing
peptide-pulsed targets. This suggests that T-cell based vaccines would
provide sterilizing immunity if memory CD8$^+$T cells, that are generated
by vaccination, were to remain present at high enough frequencies
\cite{Schmidt.pnas08,Hansen.nm09}.

This analysis may form a basis for quantitative understanding of
efficacy of T cell-based vaccines.  By correlating expression of
various cell surface and intracellular markers with the \vivo
killing efficacy of memory T cells of different specificities in mice,
we may better understand which qualities of memory cells provide best
protection. For example, one could ask if polyfunctional CD8$^+$T cell
induced by vaccination are better killers \vivo than monofunctional
memory T cells \cite{harari.ir06}. Such information can potentially be
further used to predict efficacy of T cell-based vaccines in humans.

\section{Material and Methods}

\subsection{Cytotoxicity \vivo}

Experimental method of measuring cytotoxicity of CD8$^+$ T cells \vivo
has been describes in great detail elsewhere (e.g.,
\cite{Ingulli.mmb07}).  In this report, we analyze recently published
data on killing of peptide-pulsed splenocytes by LCMV-specific
effector and memory CD8$^+$ T cells \cite[\& see
\fref{cartoon}]{Barber.ji03}.  The reader is referred to the original
publication for more detail.  In the first set of experiments (``in
vivo LCMV infection''), target splenocytes were pulsed with NP396 or
GP276 peptides of LCMV ($10\ \mu$M) or left unpulsed.  Targets were
subsequently transferred into syngenic mice either infected with LCMV
8 days previously (``acutely infected'' mice) or recovered from LCMV
infection (LCMV-immune or ``memory'' mice).  At different times after
the transfer, spleens were harvested, and the number of pulsed and
unpulsed targets, splenocytes, and peptide-specific CD8$^+$T cells was
calculated.

In the second set of experiments (``adoptive transfer''), $10^6$ of
P14 CD8$^+$T cells, expressing a TCR specific for the GP33 epitope of
LCMV, were adoptively transferred into recipient B6 mice which were
then infected i.p.  with LCMV-Arm \cite{Kaech.ni01}.  Eight (for
effectors) or 40 (for memory T cells) days later, different numbers of
P14 CD8$^+$T cells harvested from these mice were transferred into new
naive recipients (\fref{cartoon}B).  The number of effector CD8$^+$T
cells transferred into different recipients was $10^6$, $2\times10^6$,
$10^7$, and $2\times10^7$. The number of memory CD8$^+$T cells
transferred into different recipients was $10^6$, $2\times10^6$, and
$10^7$. Two hours later, two populations of CFSE labeled splenocytes,
one of which was pulsed with the GP33 peptide of LCMV (1 $\mu$M), were
transferred into these recipient mice, harboring the transferred
GP33-specific effector or memory CD8$^+$T cells. Percent targets killed
was calculated at different times after target cell transfer as
described earlier \cite{Barber.ji03,Ingulli.mmb07}. The ratio of the
frequency of pulsed to unpulsed targets, used in fitting of the data,
was calculated as $R=1-L/100$ where $L$ is the percent of
peptide-pulsed targets killed \cite{Ingulli.mmb07,Ganusov.jv08}.

\begin{figure}
\begin{center}
  \includegraphics[width=.95\textwidth]{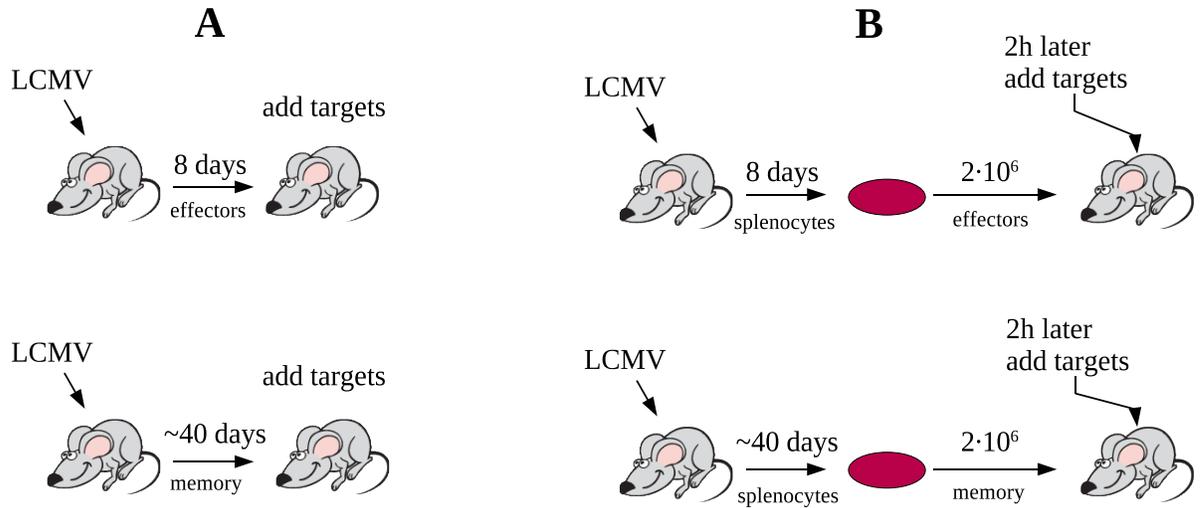}
\end{center}
\caption{Schematic representation of the \vivo cytotoxicity assays
  undertaken to investigate the quantitative details of CD8$^+$T cell
  mediated killing of peptide-pulsed targets in the mouse spleen. In
  the first set of experiments (``LCMV infection'', panel A), B6 mice
  were infected with LCMV-Arm and 8 or 37-100 days later, three
  populations of $5\times10^6$ target cells (pulsed with either NP296
  or GP276 peptides of LCMV and unpulsed) were transferred into these
  mice.  In the second set of experiments (``adoptive transfer'',
  panel B), P14 TCR Tg CD8$^+$T cells, specific to the GP33 epitope of
  LCMV, were transferred into B6 mice and then infected with LCMV-Arm.
  Eight or 40 days later, different number of effector (day 8) or
  memory (day 40) P14 CD8$^+$T cells from these mice were transferred
  into new naive B6 mice. In panel B, we shown an example of
  $2\times10^6$ effectors or memory CD8$^+$T cells transferred.  Two
  hours later, two populations of $5\times10^6$ targets (pulsed with
  the GP33 peptide of LCMV and unpulsed) were transferred into these
  mice now harboring GP33-specific CD8$^+$T cells. In both sets of
  experiments, killing of peptide-pulsed targets was measured in
  spleens of mice at different times after cell transfer
  \cite{Barber.ji03}.
}\label{fig:cartoon}
\end{figure}

\subsection{Mathematical model for the cytotoxicity \vivo assay}

Details of the mathematical model proposed to describe migration of
injected targets from the blood to the spleen and killing of
peptide-pulsed targets in the spleen are given in great detail
elsewhere \cite[see also Supplementary Information]{Ganusov.jv08}.  In
short, target cells injected i.v.  migrate from the blood to the
spleen at a rate $\sigma$, die at a rate $\eps$ due to preparation
techniques (independent of CD8$^+$T cell mediated killing), or migrate to
other tissues and/or die elsewhere at a rate $\delta$.  In the spleen,
targets die due to preparation-induced death rate $\eps$, and
peptide-pulsed targets also die due to CD8$^+$T cell mediated killing,
described by the rate $K$. The dynamics of unpulsed targets $S(t)$ and
the ratio of the frequency of peptide-pulsed to unpulsed targets
$R(t)$ in the spleen is given by equations \cite{Ganusov.jv08}

\beqa S(t) &=& {S_B(0)\sigma \over d-\eps}\left[1-{\rm
    e}^{-(d-\eps)t}\right]{\rm e}^{-\eps t},
\label{eqn:S} \\
R(t) &=& {(d-\eps)\over (K-(d-\eps))}\left[{{\rm e}^{-(d-\eps)t}-{\rm
      e}^{-Kt}\over 1-{\rm e}^{-(d-\eps) t}} \right]{\rm e}^{-\eps
  t},\label{eqn:R}
\eea

\no where $d=\sigma+\eps+\delta$ is the rate of removal of cells from
the blood and $S_B(0)=5\times10^6$ is the initial number of unpulsed
targets in the blood \cite{Barber.ji03}.

We have shown previously that the rate of recruitment of target cells
from the blood to the spleen depends on the spleen size
\cite{Ganusov.jv08}. Therefore, to describe recruitment of targets
into the spleen we let the rate of recruitment be $\sigma = \al\times
{N_s}_i$ where ${N_s}_i$ is the number of splenocytes in the $i^{th}$
mouse and $\al$ is a coefficient \cite{Ganusov.jv08}.

In our previous study we estimated the death rate of targets $K$,
pulsed with NP396 or GP275 peptides of LCMV, due to killing by the
total effector or memory CD8$^+$T cell response \cite{Ganusov.jv08}. To
estimate the per capita killing efficacy of CD8$^+$T cells, we have to
relate the death rate of peptide pulsed targets $K$ to the density of
epitope-specific CD8$^+$T cells in the mouse spleen. It is generally
assumed that killing of targets follows the law of mass action, that
is the death rate of peptide-pulsed targets is proportional to the
frequency of peptide-specific CD8$^+$T cells
\cite{Regoes.pnas07,Yates.po07}.  However, spleen tissue has a
complicated structure and the assumption of mass-action like encounter
of targets and CD8$^+$T cells need not hold.  Therefore, here we test
several different killing terms in how well they describe the data
from the \vivo cytotoxicity assay. In a mass-action model, killing
occurs at a rate that is proportional to the frequency of targets and
the frequency of epitope-specific CD8$^+$T cells $E_i$ in the spleen of
the $i^{th}$ mouse, i.e., the death rate of peptide-pulsed targets in
the $i^{th}$ mouse due to CD8$^+$T cell mediated killing is

\beq K=kE_i\label{eqn:mass-action}\ee

Alternatively, it is possible that the encounter rate between targets
and killers does not follow the law of mass-action and is affected by
the frequency or the number of CD8$^+$T cells and/or the number of
frequency of targets. For example, the death rate of peptide-pulsed
targets may saturate with increasing killer frequencies of decrease
with increasing target cell frequencies (see Supplementary Information
for more detail).

To fit the data on recruitment of targets into the spleen and on
killing of peptide-pulsed targets in the spleen at the same time we
$\log$-transform the data and the model predictions.  To access lack
of fit of the data with repeated measurements we use the F-test
\cite[p.  29]{Bates.b88}. To compare nested models we also use the
F-test \cite[p.  104]{Bates.b88}.  Fittings were done in Mathematica
5.2 using the routine FindMinimum.

\section{Results}

\subsection{Killing efficacy of T cells following acute LCMV
  infection}

To investigate quantitative aspects of how effector and memory CD8$^+$T
cells kill their targets in a mouse spleen we analyze data from
recently published experiments on killing of targets pulsed with
either NP396, GP276, or GP33 peptides from LCMV by peptide-specific
effector or memory CD8$^+$T cells \cite{Barber.ji03}. Mice, infected with
LCMV-Armstrong develop a vigorous CD8$^+$T cell response that peaks 8
days after the infection \cite{Murali-Krishna.i98,Homann.nm01}. By
15-30 days after the infection, most of effectors die and a population
of LCMV-specific memory CD8$^+$T cells persists for the life of the
animal \cite{Homann.nm01}. To measure the efficacy of effector and
memory CD8$^+$T cells, target cells pulsed with LCMV-specific peptides
(NP396 or GP276) were transferred into mice infected 8 or $37-100$
days previously \cite{Barber.ji03}, and the percent of targets killed
by effectors or memory CD8$^+$T cells was calculated (see Materials and
Methods and \fref{cartoon}).

We have previously developed a mathematical model to estimate the
killing efficacy of LCMV-specific T cell responses from the data
obtained in such \vivo cytotoxicity assay \cite{Ganusov.jv08}. The
model describes the most important processes: recruitment of target
cells from the blood to the spleen, death of targets due to
preparation, and killing of peptide-pulsed targets by the total
peptide-specific CD8$^+$T cell response in the spleen \cite[see Materials
and Methods]{Ganusov.jv08}.  From these data we estimated the death
rate of peptide-pulsed targets due to killing by the total effector or
memory CD8$^+$T cell response \cite{Ganusov.jv08}.

Here we extend this model by allowing different terms for the killing
of targets by epitope-specific CD8$^+$T cells (see Materials and
Methods).  Using the approach of a previous study
\cite{Regoes.pnas07}, we first fitted the data from acutely infected
and LCMV-immune mice assuming that killing of targets is proportional
to the frequency of epitope-specific CD8$^+$T cells in the spleen of a
given mouse.  Although this model fits the data, the quality of the
fit to the data was rather poor (lack of fit test: $F_{30,158}=3.1$,
$p=2.6\times10^{-6}$).  Assuming that the death rate of peptide-pulsed
targets $K$ is dependent on the total number of epitope-specific CD8$^+$T
cells, rather than their frequency, led to even worse fits of the data
(lack of fit test: $F_{30,158}=5.54$, $p=3.4\times10^{-13}$).  The
reason for the poor fit is that the frequency (or total number) of
epitope-specific CD8$^+$T cells measured in an individual mouse predicted
poorly killing of targets in the same mouse
(\fref{prediction-ind-killing} in Supplementary Information).

We therefore explored several modifications of the model to improve
the fit of the model to the data (see Supplementary Information). The
best description of the data was obtained by assuming that killing of
targets is determined by the average frequency of NP396- and
GP276-specific effector or memory CD8$^+$T cells in the spleen and not
values measured in individual mice (lack of fit test:
$F_{30,162}=0.79$, $p=0.77$; see \tref{parameters} and \fref{fits} in
Supplementary Information). The good fit of this model to data
suggests that variation in the frequency of epitope-specific CD8$^+$T
cells measured in different mice is largely due to measurement noise.
The absence of a positive correlation between the number of targets
killed and the CD8$^+$T cell frequency in a given mouse further supports
this conclusion (\fref{ratio_vs_cd8}).  This analysis suggested that
LCMV-specific CD8$^+$T cells are half as efficient as are effector CD8$^+$T
cells of the same specificity (\tref{parameters}).

Another version of the model in which the death rate of targets due to
CD8$^+$T cell killing saturates with increasing T cell frequency (see
\eref{saturation_E} in Supplementary Information) significantly
improved the fit of the model to data (F-test for nested models:
$F_{1,186}=50.3$, $p=2.7\times10^{-11}$).  This model predicted that
memory CD8$^+$T cells are 10 fold less efficient killers than effector T
cells of the same specificity (see Supplemental Information). It
should be noted, however, that if measurements of CD8$^+$T cell
frequencies in individual mice are noisy, saturation in killing with T
cell frequency is expected to be important since it allows for
smoothing of noisy data.

\begin{table}
\small
\bc
\begin{tabular}{|l|cl||l|c|c|c|l|}
  \hline
  Parameter & Mean & 95\% CIs &$E/\freq$&$E$, \% & $E$, $10^6$ cells&
  cell type\\
  \hline
  $\al_A,\ 10^{-12}\ min^{-1}$&$7.17$&$5.63-9.25$&&&&\\
  $\al_M,\ 10^{-11}\ min^{-1}$&$1.30$&$0.90-1.88$&&&&\\
  $\eps,\ 10^{-3}\ min^{-1}$&$4.71$&$3.37-6.12$&&&&\\
  $\gamma$&0.49&$0.35-0.66$ &&&&\\
  $k_{NP396},\ min^{-1}$&5.50&$4.26-7.21$&2471&6.3&10.4&NP396-spec effectors\\
  $k_{GP276},\ min^{-1}$&2.35&$1.85-2.89$ &130&2.1&3.6&GP276-spec effectors\\
  $\gamma k_{NP396},\ min^{-1}$&2.68&$2.10-3.32$ &4.9&0.54&0.32&NP396-spec memory\\
  $\gamma k_{GP276},\ min^{-1}$&1.14&$0.79-1.54$
  &1.5&0.35&0.21&GP276-spec memory\\
\hline
\end{tabular}
\end{center}
\caption{Parameters providing the best fit of the mathematical model
  assuming that the rate of recruitment of targets into the spleen
  depends on the spleen size and that killing of peptide-pulsed
  targets depends on the average frequency of epitope-specific CD8$^+$T
  cells in the spleen.  Here $\al_A$ and $\al_M$ are coefficients
  relating the recruitment rate of cells into the spleen $\sigma=\al
  {N_s}_i$ in acutely infected ($\al_A$) and memory ($\al_M$) mice,
  and ${N_s}_i$ is the number of splenocytes in individual mice,
  $\gamma$ is the ratio of the killing efficacy of epitope-specific
  memory CD8$^+$T cells to that of effector CD8$^+$T cells, $k_{NP396}$ and
  $k_{GP276}$ are the per capita killing efficacy of NP396- and
  GP276-specific effector CD8$^+$T cells, and $\gamma k_{NP396}$ and
  $\gamma k_{GP276}$ are the killing efficacy of NP396- and
  GP276-specific memory CD8$^+$T cells, respectively.  In the fits the
  rate of migration of labeled splenocytes to other organs $\delta$
  was fixed to 0 since this did not affect the quality of the model
  fit to data (F-test for nested models: $F_{1,191}=0.15$, $p=0.70$).
  Data and model fits are shown in \fref{fits}. CIs were calculated by
  bootstrapping the data with 1000 simulations \cite{Efron.b93}. Note
  that the fits predict that epitope-specific memory CD8$^+$T cells on
  the per capita basis are half as efficient as effectors
  ($\gamma=\gamma_{NP396}= \gamma_{GP276}\approx 0.5$; F-test for
  nested models: $F_{1,191}=0.03$, $p=0.88$). For different
  experiments, we also show the average effector to target ratio
  ($E/\freq$), the average percentage, and the average total number of
  epitope-specific CD8$^+$T in the spleen obtained from the data.
}\label{tab:parameters}
\end{table}

\subsection{Killing efficacy of T cells following adoptive transfer of
  T cells}

Thus, we generated two alternative models that provided a good
description of the data but generated highly distinct predictions on
the efficiency and nature of killing of targets in the mouse spleen.
To discriminate between these alternative models we analyzed novel
data from additional experiments involving transfer of different
numbers of effector or memory CD8$^+$T cells specific to the GP33 epitope
of LCMV \cite[see \fref{cartoon}B]{Barber.ji03}. Two hours after
transfer of CD8$^+$T cells, GP33-pulsed and unpulsed target cells were
transferred into mice harboring GP33-specific CD8$^+$T cells, and killing
of peptide-pulsed targets was measured longitudinally
\cite[\fref{fits-transfer}]{Barber.ji03}.  Approximately 2 to 10\% of
the adoptively transferred CD8$^+$T cells accumulated in the mouse spleen
(\tref{parameters-transfer}). Since the transfer of different numbers
of epitope-specific CD8$^+$T cells led to different frequencies of these
cells in the spleen as well as to different effector to target ratios
(\tref{parameters-transfer}), these data allowed for a unique
opportunity to investigate whether the per capita killing efficacy of
LCMV-specific effector and memory CD8$^+$T cells is independent of these
two quantities.

Therefore, we fitted the mathematical model given in
\eref{S}-(\ref{eqn:R}) to these data assuming that the death rate of
peptide-pulsed targets $K$ depends on the average frequency of
GP33-specific CD8$^+$T cells in the spleen (i.e., killing follows the law
of mass action). The model described the data very well with the
exception of one time point with very few unpulsed targets being
recruited into the spleen (\fref{fits-transfer}A at $2\times10^7$
effector CD8$^+$T cell transferred; lack of fit test with this time point
removed: $F_{20,50}=0.92$, $p=0.56$). By fitting the model, we
estimated parameters determining the rate of migration of targets from
the blood to the spleen as well as the per capita killing efficacy of
GP33-specific effector (given by $k_i$) and memory (given
$\gamma_i\times k_i$) CD8$^+$T cells, at different frequencies of
effectors of memory CD8$^+$T cells in the spleen
(\tref{parameters-transfer} and \fref{killing}).  Surprisingly, the
model fits predicted that the per capita killing efficacy of effector
CD8$^+$T cells was largely independent of the frequency of effectors in
the spleen equaling on average $k_{GP33} = 2.1\pm0.17\ min^{-1}$. Most
importantly, this estimate is almost identical to the estimate of the
killing efficacy of GP276-specific effectors obtained above by fitting
the data from acute LCMV infection using the average frequency of
epitope-specific CD8$^+$T cells in the spleen (see \tref{parameters}).
Thus, this analysis suggests that changing the frequency of
epitope-specific CD8$^+$T cells in the mouse spleen from $6\times10^{-4}$
(transfer of $10^6$ GP33-specific effectors, see
\tref{parameters-transfer}) to $2\times10^{-2}$ (GP276-) or
$6\times10^{-2}$ (NP396-specific effectors, see \tref{parameters})
does not affect the per capita killing efficacy of LCMV-specific
effector CD8$^+$T cells. In other words, this implies that killing of
targets in the mouse spleen by effector CD8$^+$T cells follows the law of
mass action where the rate of killing is proportional to the frequency
of epitope-specific CD8$^+$T cells. This conclusion was further confirmed
by the lack of improvement of the data fit with the model assuming
saturation in the death rate of peptide-pulsed targets with CTL
frequency (F-test for nested models: $F_{1,16}=0.62$, $p=0.44$).
Thus, for LCMV-specific effectors, the correct model is the one in
which killing of targets is simply proportional to the average
frequency of CTLs in the spleen.

In contrast with effectors, we estimated that the per capita killing
efficacy of memory CD8$^+$T cells declines with an increasing number of
memory T cell transferred (\tref{parameters-transfer} and
\fref{killing}).  The killing efficacy of memory CD8$^+$T cells was
somewhat higher than that of effectors at the lowest number of
epitope-specific CD8$^+$T cell transferred, and lower when larger numbers
of memory T cells were transferred (\fref{killing}). This suggests
that the death rate of peptide-pulsed targets due to killing by memory
CD8$^+$T cells saturates as the function of the frequency of CD8$^+$T cells
(or the effector to target ratio) in the mouse spleen. Including
saturation in the death rate of targets with memory CD8$^+$T cell
frequency (see \eref{saturation_E}) significantly improved the quality
of the model fit to data on killing of peptide-pulsed targets by
NP396- and GP276-specific memory CD8$^+$T cells (F-test for nested
models: $F_{1,78}=15.3$, $p=2.0\times10^{-4}$).  These fits also
predicted that at low T cell frequencies, memory T cells are at least
as efficient as effectors of the same specificity (for memory CD8$^+$T
cells assuming saturation in killing: $k_{NP396}=6.2\ min^{-1}$,
$k_{GP276}=2.2\ min^{-1}$, $c_E=302.4$; compare these estimates to the
killing efficacy of effectors given in \tref{parameters}), confirming
the result from the adoptive transfer experiment. Moreover, the
estimated saturation constant $c_E$ implies that killing efficacy of
memory T cells is reduced by half at the cell frequency equal
$1/c_E\approx 0.35\%$ which is again observed in both following LCMV
infection (\tref{parameters}) and after adoptive transfer of
GP33-specific memory T cells (\tref{parameters-transfer}).  Thus,
these results suggest that the death rate of peptide-pulsed targets
saturates with increasing the frequency of memory CD8$^+$T cells in the
mouse spleen.

Assuming that the death rate of peptide-pulsed targets is proportional
to the total number of epitope-specific CD8$^+$T cells in the mouse
spleen, we found that following acute LCMV infection effector and
memory CD8$^+$T cells have a similar per capita killing efficacy (results
not shown). This is contrast to the result obtained from the adoptive
transfer experiments where at frequencies of epitope-specific effector
and memory CD8$^+$T cells in the spleen of $\sim 0.35\%$, memory T cells
are only 30\% to 50\% as efficient as effectors. This suggests that
killing targets in the spleen is proportional to the frequency, and
not the total number of epitope-specific CD8$^+$T cells confirming a
previously made assumption \cite{Regoes.pnas07,Yates.po07}.

\begin{figure}
\begin{center}
  \includegraphics[width=.99\textwidth]
{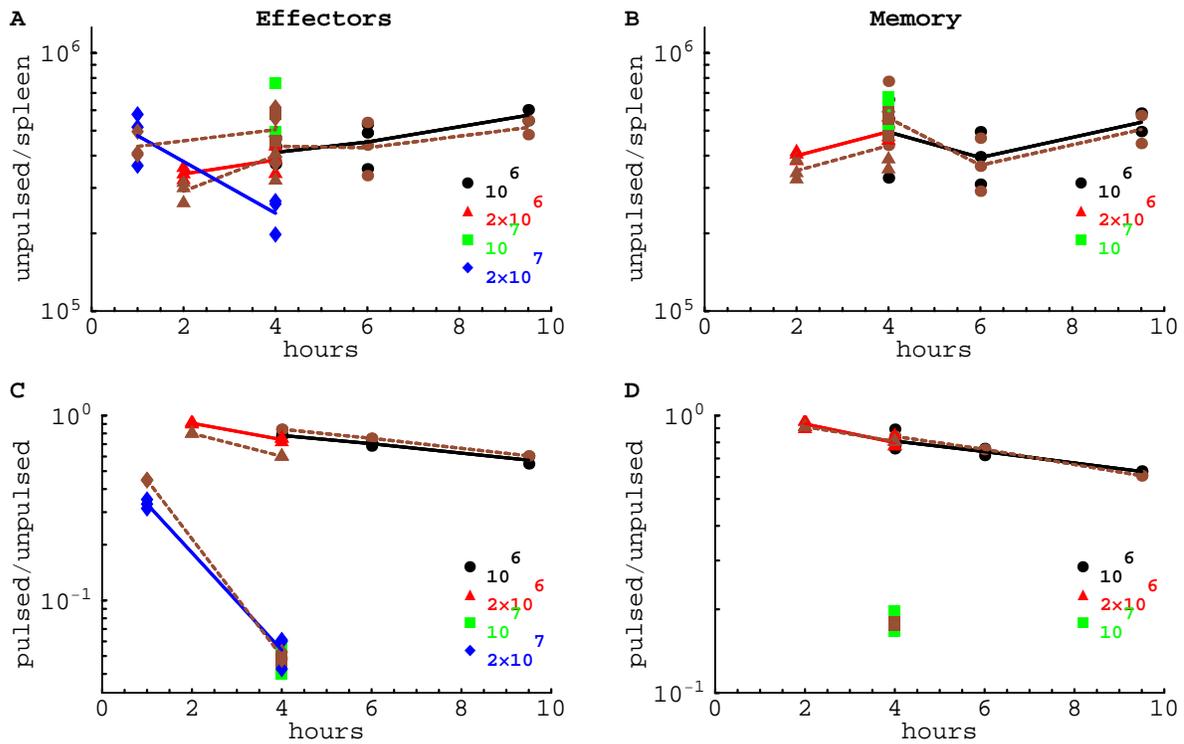}
\end{center}
\caption{Fits of the mathematical model to data from  experiments
involving adoptive transfer of different numbers of epitope-specific
effector (panels A\&C) or memory (panels B\&D) CD8$^+$T cells. Panels A
and B show the number of unpulsed targets in the spleen  
at different times after
  cell transfer. Panels C and D show the change in the ratio of the
  frequency of peptide-pulsed to unpulsed targets in the spleen with time.
  Different symbols denote data from different adoptive transfer
  experiments with $10^6$, $2\times10^6$, $10^7$, or $2\times10^7$ T
  cells transferred. Symbols denote individual measurements with
  averages per time point being connected by solid lines. Brown
  symbols are the model predictions with averages being connected by
  dashed lines.  Parameters providing the best fits of the model are
  shown in \tref{parameters-transfer}.  Note that in panel A, the
  model does not predict the decline in the number of unpulsed targets
  with time in experiments with transfer of $2\times10^7$
  GP33-specific effectors. Such decline in the number of unpulsed
  targets in the spleen is unexpected and is most likely due to a
  measurement error.
}\label{fig:fits-transfer}
\end{figure}

\begin{table}
\bc
\begin{tabular}{|l|ll||c|c|c|c|}\hline
  Parameter&Mean&95\% CIs&$E/\freq$&$E$, \% & $E$, $10^6$ cells &Cells transferred\\
  \hline
  $\al,\ 10^{-11}\ min^{-1}\ cell^{-1}$&2.14&1.88--6.16&&&&\\
  $\eps,\ 10^{-3}\ min^{-1}$&1.1&0.4--1.6&&&&\\
  $\delta,\ 10^{-2}\ min^{-1}$&1.0&0.7--4.3&&&&\\
  \hline
  $k_1,\ min^{-1}$&1.77&0.91--2.12&0.14&0.06&0.05&\multirow{2}{*}{$10^6$}\\
  $\gamma_1k_1,\ min^{-1}$&3.26&2.54--3.83&0.09&0.04&0.03&\\ \hline
  $k_2,\ min^{-1}$&1.77&0.91--2.12&0.34&0.18&0.10&\multirow{2}{*}{$2\times10^6$}\\
  $\gamma_2k_2,\ min^{-1}$&0.98&0.54--1.22&0.25&0.15&0.09&\\
  \hline
  $k_3,\ min^{-1}$&3.08&1.6--3.9&23.7&0.87&0.68&\multirow{2}{*}{$10^7$}\\
  $\gamma_3k_3,\ min^{-1}$&0.93&0.61--1.07&9.48&1.25&1.05&\\
  \hline
  $k_4,\ min^{-1}$&1.77&0.91--2.12&44.4&1.54&1.43&$2\times10^7$\\
  \hline
\end{tabular}
\end{center}
\caption{Estimates of parameters of the mathematical model fitted to
  the data from the adoptive transfer experiments. In different
  experiments, $10^6$, $2\times10^6$, $10^7$ or $2\times 10^7$
  effector or memory CD8$^+$T cells were transferred resulting in the
  shown average effector to target ratio $E/\freq$, average percentage
  or the total number of transferred cells in the recipient mice.  We
  estimated the killing efficacy of effectors ($k_i$) and the ratio of
  the killing efficacy of an effector to that of a memory cell
  ($\gamma_i$) by assuming that the death rate of peptide pulsed
  targets is proportional to the frequency of epitope-specific CD8$^+$T
  cells in the spleen, $K=kE$ where $k$ and $\gamma k$ is the killing
  efficacy of GP33-specific effector and memory CD8$^+$T cells,
  respectively.  Killing efficacies $k_1$, $k_2$ and $k_4$ were fitted
  as one parameter since this did not significantly affect the quality
  of the model fit to data ($F_{2,68}=2.84$, $p=0.07$).  Further
  reduction of the number of model parameters resulted in the
  significantly worse description of the data (results not shown).
  Interestingly, this data also lead to a non-zero estimate of the
  preparation-induced cell death rate that we have previously
  postulated to exist \cite{Ganusov.jv08}, although in these
  experiments this rate was smaller than during acute LCMV infection
  (see \tref{parameters}).  }\label{tab:parameters-transfer}
\end{table}

\begin{figure}
\begin{center}
  \includegraphics[width=.99\textwidth]%
  {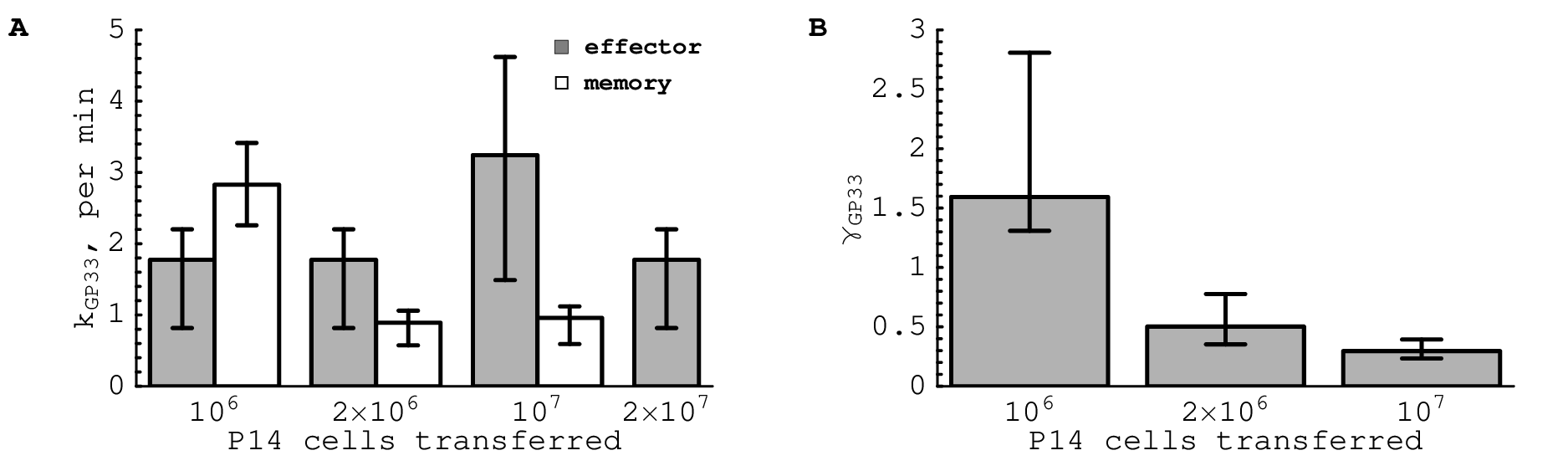}
\end{center}
\caption{Estimated per capita killing efficacy of GP33-specific
  effector CD8$^+$T cells (A) and the ratio of killing efficacy of
  GP33-specific memory to that of effector CD8$^+$T cells (B) as the
  function of the number of transferred effectors or memory cells.
  Error bars show the 95\% confidence intervals for estimated
  parameters. These results suggest that there is a minimal change in
  the per capita killing efficacy of GP33-specific effector CD8$^+$T
  cells with effector T cell frequency (with an average of
  $k_{GP33}=2.1\pm0.17\ min^{-1}$) but that the efficacy of memory CD8
  T cells declines at high numbers of transferred cells.
}\label{fig:killing}
\end{figure}

\section{Discussion}

Recent interest in T cell based vaccines against several chronic
infections of humans requires the development of experimental and
theoretical tools to access the efficacy of such vaccines
\cite{Sekaly.jem08,DeBoer.jv07}.  It is generally believed that memory
CD8$^+$T cells induced by vaccination are not able to provide sterilizing
immunity, because T cells react only to infected cells, i.e., after
the infection has been established. However, a recent study has shown
an example where generation of a large population of memory CD8$^+$T
cells by vaccination in mice did provide sterilizing immunity against
malaria \cite{Schmidt.pnas08}.

Quantitative approaches aimed at estimating the \vivo efficacy of
effector and memory CD8$^+$T cells, and at quantitative details of how
CD8$^+$T cells control growth of pathogens are necessary to understand
how protection induced by T cell-based vaccines is achieved. This
study utilized a recently developed experimental technique of \vivo
cytotoxicity to investigate how effector and memory CD8$^+$T cells,
specific to LCMV, kill peptide-pulsed targets in the mouse spleen.
Using a novel mathematical model we have analyzed data from
experiments on \vivo killing of targets following LCMV infection of
mice and experiments involving adoptive transfer of different numbers
of LCMV-specific effector or memory CD8$^+$T cells. Results of our
analysis suggest that death rate of targets due to LCMV-specific
effectors is simply proportional to the average frequency of
epitope-specific CD8$^+$T cells in the mouse spleen. This is a rather
surprising result.  Even for frequencies of LCMV-specific effectors in
the spleen ranging over 100 fold from 0.06\% to 6\%, we find no
evidence in saturation in the death rate of peptide-pulsed targets
with CD8$^+$T cell frequency. This in turn suggests that killing of
targets by CTLs follows the law of mass action
\cite{Chandrasekhar.rmp43} and that CTLs do not compete for access to
targets at least when their frequency in the spleen is as high as 6\%.

We found that killing of targets is dependent on the frequency of
epitope-specific CD8$^+$T cells in the spleen and not their total number.
This has important implications for vaccination since this suggests
that inducing high numbers of virus-specific CD8$^+$T cells may not be
highly advantageous if their frequency in tissues is low.  The
observation that the death rate of peptide-pulsed targets is mainly
determined by the average frequency of epitope-specific CTLs suggests
that a large variation in the frequency of CD8$^+$T cells in the spleen,
as measured in individual mice, represents measurement noise.

We found that at low CD8$^+$T cell frequencies, memory CD8$^+$T cells
are at least as efficient at killing peptide-pulsed targets as are
effector CD8$^+$T cells. This contradicts the widely accepted view that
memory CD8$^+$T cells, especially those residing in lymphoid tissues, are
not very efficient killers and generally require restimulation to
exhibit cytotoxicity \cite{Masopust.s01}. Our study as well as the
original work of \citet{Barber.ji03} illustrates the limitation of the
in vitro chromium release assay to assess cytotoxic efficacy of memory
CD8$^+$T cells.

We also found that at high frequencies (e.g., following acute LCMV
infection), LCMV-specific memory T cells are only half as efficient as
are effector CD8$^+$T cells of the same specificity.  The reduction in
the killing efficacy of memory CD8$^+$T cells with increasing their
frequency may limit the overall efficacy of T cell-based vaccines,
since boosting the frequency of memory T cells may not lead to a
proportional increase in the efficacy of the memory T cell response.
Different efficacies of effector and memory CD8$^+$T cells following LCMV
infection may be due to different localization of these two subsets in
the spleen. A recent study suggests that LCMV-specific effectors
localize mainly in the red pulp, while memory CD8$^+$T cells reside in
the T cell zones of the white pulp of the mouse spleen
\cite{Dauner.ji08}. Since white pulp occupies the minority of the
space in the spleen (5-20\%, \cite{Freitas.b03}), memory CD8$^+$T cells
indeed are expected to compete for the access to targets if their
density in the white pulp is high.  However, additional studies
addressing the question of localization of target cells in the mouse
spleen are needed to investigate this issue further.

We found that NP396-specific effector CD8$^+$T cells present at the peak
of the immune response have a per capita killing efficacy
$k_{NP396}=5.5\ min^{-1}$. This value has a simple interpretation as
the death rate of targets when the frequency of CTLs in the spleen is
close to 1. For example, if most of splenocytes were NP396-specific
CD8$^+$T cells, NP396-pulsed targets would have a half-life of $\ln 2/5.5
= 7.5$ seconds in the spleen. Another interesting parameter that can
be calculated from our study is the number of targets killed by one
CTL per unit of time. In an extreme situation where all splenocytes
are targets for a single CTL a NP396-specific CTL is expected to kill
$k_{NP396}\times 1/N_s\times N_s=k_{NP396}\approx 5$ targets per
minute (or $7.2\times10^3$ targets per day). In our experiments, the
number of targets was generally much smaller than the number of
splenocytes, and since the number of targets killed per day per CTL
depends on the number of targets, we previously estimated that
following LCMV infection, CTLs kill only a few targets per day at
these high effector to target ratios \cite{Ganusov.jv08}.

The time taken by a CTL to kill its target is a sum of the time
required by a CTL to find its target and the time to deliver the
lethal hit.  By assuming that the rate of killing of peptide-pulsed
targets is limited mainly by the time required by an effector T cell
to find its target, we calculated an approximate upper bound estimate
for the killing efficacy of T cells. Given the estimated motility of
activated T cells in lymph nodes and the size of targets, the maximum
killing efficacy of CD8$^+$T cells is $k \approx 72\ min^{-1}$, although
this value does depend on the several parameters that have not been
measured in our experiments (see Supplementary Information). Since the
estimated killing efficacy of effector T cells is much lower than this
value, our result suggests that the main limiting step of killing of
peptide-pulsed targets is the actual process of killing, and not
finding the target.  This is consistent with a recent study on imaging
of killing of B cells by effector CD8$^+$T cells where the process of
killing of targets took 10 to 20 minutes \cite{Mempel.i06b}. The
observation that killing efficacy of effector CD8$^+$T cells is the same
at low and high effector frequencies in the spleen also suggests that
finding the target is not the limiting step in the killing process.

Effector CD8$^+$T cells, specific to two other epitopes of LCMV, are
estimated to have a lower killing efficacy ($k_{GP276}=2.4\ min^{-1}$,
$k_{GP33}=2.1\ min^{-1}$). Since we do not expect these cells to
behave differently from NP396-specific effectors, a lower killing
efficacy of GP276- and GP33-specific CD8$^+$T cells is likely because of
longer times required by effectors to kill their targets. This may
arise because of the reduced affinity of T cell receptors, specific to
these peptides, to their ligands or a shorter half-life time of
peptide-MHC complexes on target cells.

Reduced killing efficacy of memory CD8$^+$T cells as compared to
effectors of the same specificity is expected as memory T cells
generally express low levels of molecules such as perforin, granzymes,
and FasL that are required for mediating cytotoxicity
\cite{Barber.ji03}. Even though memory CD8$^+$T cells may have a reduced
motility in lymphoid tissues as compared to activated, effector T
cells, this is an unlikely reason for their reduced killing efficacy.
We have found that at low T cell frequencies GP33-specific memory CD8
T cells are at least as efficient killers as are effectors. This may
imply that high levels of granzymes and perforin that are expressed by
effector CD8$^+$T cells are not required for high degree of cytotoxicity
\vivo.

Our results have important implication for the loss of protection by
memory CD8$^+$T cells. It has been suggested that infection with new
pathogens leads to attrition (loss) of memory CD8$^+$T cells specific to
previously encountered pathogens
\cite{Selin.i99,Antia.pnas98,Ganusov.pcb06,Welsh.n09}. A recent study
has challenged this conclusion by showing that infection of mice with
Vesicular stomatitis virus (VSV) and Vaccinia virus (VV) can also lead
to an increase in the total number of memory T cells, and as the
result, to a very moderate loss of the total number of memory CD8$^+$T
cells, specific to previously encountered virus (LCMV), in the spleen
\cite{Vezys.n08}. However, this study has also shown a dramatic
reduction in the frequency of LCMV-specific memory CD8$^+$T cells in the
spleen and in peripheral tissues following infection with VSV and VV.
Our results suggest that reduction in the frequency of virus-specific
memory CD8$^+$T cells may have a dramatic affect on the efficacy of the
memory response even if the total number of virus-specific CD8$^+$T cells
is not dramatically reduced.  Reduction in the cytotoxic efficacy of
Vaccinia virus- or Pichindie virus-specific memory CD8$^+$T cell
responses has indeed been observed after exposure to LCMV
\cite{Kim.ji04,Welsh.n09}.

Our results suggest a simple procedure of estimating the frequency of
memory CD8$^+$T cells at which T cells may provide sterilizing immunity
upon re-exposure to a virus. From a simple equation governing the
growth of the virus $V'(t) = (r-\frac{kE}{1+c_E E})V(t)$, this
frequency of memory CD8$^+$T cells is given by the ratio
$E=\frac{r}{k-rc_E}$ where $r$ is the rate of replication of the
virus, $k$ is the killing efficacy of memory CD8$^+$T cells, and $c_E$ is
the half-saturation constant. For a virus such as LCMV that doubles
its population size in $\ln2/5 = 3.4$ hours
\cite{Ehl.eji97,Bocharov.jv04,Althaus.ji07}, the protective frequency
of memory CD8$^+$T cells which have a killing efficacy $k=2.5\ min^{-1}$
is $5/(2.5-300\times5)\approx 0.24\%$ of all splenocytes.  This is
below the level of memory CD8$^+$T cells induced by vaccination with
LCMV-Armstrong (see \tref{parameters}), suggesting that LCMV-immune
mice should be protected following infection with other strains of
LCMV even in the absence of LCMV-specific antibodies.  Future studies
will be necessary to address this quantitative prediction.

Our study may provide practical guidelines for estimating the efficacy
of T cell based vaccines. It has been recently shown that cytotoxic
potential of HIV-specific CD8$^+$T cells may be one of the most important
components of effective control of viral growth \cite{Migueles.i08}.
Because of practical difficulties in performing \vivo cytotoxicity
assay in humans, it would be interesting to correlate the \vivo
killing efficacy of murine CD8$^+$T cells with their phenotype measured
ex vivo by flow cytometry (e.g., cell surface and intra-cellular
markers).  This may allow for quantitative understanding of what
constitutes an effective memory T cell; this understanding can be
further applied to compare efficacy of T cell-based vaccines in
humans.

In our analysis, we used the ratio of the frequency of peptide-pulsed
to unpulsed targets as a measure of killing. It has been shown
previously that the ratio is indeed a less biased estimator than, for
example, the frequency of peptide-pulsed targets in the spleen
\cite{Yates.po07}. The use of the ratio, however, precludes the
analysis of whether killing of targets depends on the total number of
targets or their frequency in the spleen. Addressing this question
will be explored elsewhere. We have investigated quantitative aspects
of killing of peptide-pulsed targets in the spleen. It would be
important to investigate if killing of virally infected cells and/or
targets in other organs such as the lung or the gut follows the same
principle.  Including these processes may require the use of more
sophisticated mathematical models, and as such will hopefully lead to
more collaborations between experimentalists and theoreticians.

\section*{Acknowledgments}

We thank Joost Beltman, John Wherry, Andrew Yates, Anton Zilman, Ruy
Ribeiro and Alan Perelson for comments and suggestions during this
work. This work was supported by the VICI grant 016.048.603 from NWO,
Marie Curie Incoming International Fellowship (FP6), and the U.S.
Department of Energy through the LANL/LDRD Program.

\bibliography{/home/fly10/vitaly/refs/bibliography/library-main}

\newpage
\listoffigures

\newpage
\section{Supplementary Information}

\setcounter{equation}{0}
\renewcommand{\theequation}{A.\arabic{equation}}

\subsection{Mathematical model for estimating the killing efficacy of CD8$^+$T
  cell responses}

The dynamics of unpulsed and peptide-pulsed targets in the blood and
in the spleen are given by equations
\beqa
\Dt{S_B(t)} & = & -(\delta+\sigma+\eps)S_B(t), \label{eqn:S-blood-1} \\
\Dt{S(t)} & = & \sigma S_B(t) - \eps S(t), \label{eqn:S-blood-2} \\
\Dt{T_B(t)} & = & -(\delta+\sigma+\eps) T_B(t), \label{eqn:T-blood-1} \\
\Dt{T(t)} & = & \sigma T_B(t) - \eps T(t) - K T(t), \label{eqn:T-blood-2}
\eea

\no where $S_B(t)$ and $T_B(t)$ are the numbers of unpulsed and
peptide-pulsed target cells in the blood, respectively, and $S(t)$ and
$T(t)$ is the number of unpulsed and pulsed targets in the spleen,
respectively, $\sigma$ is the rate of migration of target cells from
the blood into the spleen, and $\delta$ is the rate of cell
migration/death from blood to other organs, $\eps$ is the extra death
rate of transferred splenocytes due to preparation (independent of
epitope-specific CD8$^+$T cells), and $K$ is the death rate of
peptide-pulsed targets due to CD8$^+$T cell mediated killing in the
spleen. For our experiments, the initial conditions for the model are
$S_B(0)=T_B(0)=5\times10^6$ cells and $S(0)=T(0)=0$
\cite{Barber.ji03}.  In the case when all model parameters, including
the death rate of targets due to CD8$^+$T cell mediated killing, are
independent of time, the model can be solved analytically
\cite{Ganusov.jv08}; and the particular solution for $K=\mbox{const}$
is shown in \eref{S} and (\ref{eqn:R}).

\subsection{Deriving the general killing term}

In tissues, CD8$^+$T cells scan many cells to find virus-infected
targets, and many of the cells scanned are uninfected
\cite{Mempel.i06b}.  Scanning uninfected cells also takes some time
\cite{Mempel.i06b}, and if the majority of cells in a tissue is
uninfected, a CD8$^+$T cell can spend a substantial amount of time
``looking'' for the infected targets. The process of scanning of
uninfected targets and killing peptide-expressing targets can
described mathematically to enzyme kinetics (e.g.,
\cite{Borghans.bmb96}). We let $E$, $S$, and $T$ be the number of
killer CD8$^+$T cells, uninfected (bystander) and peptide-expressing
targets, respectively. Effector CD8$^+$T cells by scanning uninfected
targets form a complex $C_1$, and form a complex $C_2$ when they scan
infected cells. Both complexes can dissociate.  The kinetic diagram of
cell interactions is then \cite{Merrill.mb82}

\beqa
E+T&\arws{k_1}{k_{-1}}& C_1 \arw{k_2}E+D,\label{eqn:schema-1}\\
E+S&\arws{k_1}{k_{-2}}& C_2,\label{eqn:schema-2}
\eea

\no where $k_1$ and $k_{-1}$ are the rates for binding and
dissociation of a killer T cell and peptide-expressing cell; $k_{-2}$
is the dissociation rate of a complex of a killer T cell and a
bystander (uninfected) cell; $k_2$ is the dissociation rate of the
complex of killer T cell and an infected cell resulting in the death
of the infected cell (denoted as $D$). The rate of removal of pulsed
targets is then simply $k_2 C_1$. Making a quasi-steady state
assumption for $C_1$ and $C_2$, we obtain

\beqa
C_1 &=& {k_1 \over k_2 + k_{-1}} T E = K_1 T E,\label{eqn:c1-2} \\
C_2 &=& {k_1 \over k_{-2}} S E = K_2 S E. \label{eqn:c2-2} \eea

\no where $K_1 = k_1/(k_{-1}+k_2)$ and $K_2 = k_1/k_{-2}$. If the
number of killers $E$ is much larger than the number of target $T$
(i.e., $\hat E=E+C_1+C_2\approx E$, where $\hat E$ is the total number
of killer T cells), then rewriting \eref{c1-2} and (\ref{eqn:c2-2}) in
terms of the total number of unpulsed and pulsed targets, $\hat S =
S+C_2$, and $\hat T = T+C_1$, respectively, after simple algebra for
the complex $C_1$ we find

\beqa
C_1 = {K_1 E \over 1 + K_1 E}\hat T,\label{eqn:c1-3}
\eea

This suggests that the death rate of peptide-pulsed targets, $k_2
C_1/\hat T$, saturates at high numbers of killer CD8$^+$T cells
approaching the rate of dissociation of the complex $k_2$. Then the
death rate of targets is given by 

\beq K = {k E \over 1+ c_E E}, \label{eqn:saturation_E}\ee

\no where $c_E$ is the inverse frequency of CD8$^+$T cells at which
killing is half maximal.

Similarly, if the number of pulsed targets $T$ is much larger than the
number of killers $E$, then rewriting \eref{c1-2} and (\ref{eqn:c2-2})
in terms of the total number of killer CD8$^+$T cells, $\hat E =
E+C_1+C_2$, we find

\beqa C_1 = {K_1 \hat E \over 1 + K_1 T + K_2 S} T,\label{eqn:c1-4}
\eea

\no where $K_1$ and $K_2$ are defined above. This expression shows
that 1) the death rate of peptide-pulsed targets, $k_2 C_1/T$, may
decrease at high numbers of targets, and 2) if the number of bystander
(unpulsed) targets is high, $K_2 S\gg K_1 T$, the killing of targets
depends on the frequency of effectors in the spleen, $\hat E/S$, and
not on their absolute number. Simplifying \eref{c1-4} by letting
$K_2\ra 0$ we obtain

\beq K = {k E \over 1+ c_T T }, \label{eqn:saturation_T}\ee

\no where $c_T$ is the inverse frequency of pulsed targets at which
killing is half maximal.  

Finally, the death rate of a single target may depend on the effector
to target ratio, $E/T$ \cite{Abrams.tee00}. Then assuming saturation
in the death rate with the ratio of effectors to targets, we obtain

\beq K = {k (E/T)\over c_T+ (E/T)} = {k E\over E+
  c_T T}.
\label{eqn:ratio}
\ee

Note that in those cases, when the death rate of targets due to CD8$^+$T
cell mediated killing $K$ depends on the target cell density
($K=K(T)$, see \eref{saturation_T} and (\ref{eqn:ratio})), \eref{R} is
not the correct solution. Instead, we numerically solve the model
given by differential equations (see Supplementary Information) and
fit the numerical solution of the model to data.

\subsection{Deriving the killing rate constant}

Based on the assumption that the speed of a chemical reaction may be
limited by rate at which chemicals are colliding, it has been derived
from basic physical principles how the rate of reaction depends on the
properties of interacting chemicals \cite{Chandrasekhar.rmp43}.
Similar approaches have been applied in biology to model infection of
target cells by a virus \cite{Layne.pnas89,Perelson.mb93}. Bearing on
these studies, killing of peptide-pulsed target cells $T$ by
peptide-specific CD8$^+$T cells $E$ can described by a simple diagram

\beqa
E+T&\arw{k_D}& D+E,\label{eqn:kd-1}
\eea

\no with the kinetics of target cells $T$ given simply as 

\beq \Dt{T} = -k_D T E. \label{eqn:kd-2} \ee

Note that in \eref{kd-2}, $T$ and $E$ are given as cell concentrations
in the spleen (i.e., number of cells per unit of volume). From a
fundamental result of Smoluchowski \cite{Chandrasekhar.rmp43}, the
rate of the reaction $k_D$ is given by

\beq k_D = 4\pi (D_E+D_T)(R_E+R_T), \label{eqn:kd} \ee

\no where $D_E$ and $D_T$ are the diffusion (or motility) coefficients
of effectors and targets, respectively, and $R_E$ and $R_T$ are radii
of the cells.  The rate $k_D$ needs to be converted to be comparable
with the killing efficacy $k$ that we have estimated from the data
(see \tref{parameters} and \ref{tab:parameters-transfer}).  Because in
\eref{kd-2}, concentration of effectors $E$ is given as cells/volume,
converting the cell concentration to the frequency of cells in the
spleen yields
 
\beq k= {k_D N_s\over V}, \label{eqn:kD} \ee
 
\no where $N_s$ is the number of splenocytes and $V$ is the volume of
the spleen. Since spleen is packed mainly with lymphocytes, the
simplest assumption is that the spleen volume can be calculated as $v
N_s$ where $v$ is the average volume of a splenocyte given by a sphere
with the radius $R_S$. Then
 
\beq k = {4\pi (D_E+D_T)(R_E+R_T)\over v} = {3(D_E+D_T)(R_E+R_T)\over
  R_S^3}.
\label{eqn:kD-2} \ee
 
\no since volume of a splenocyte is simply $v=4/3\pi R_S^3$. Motility
(diffusion) coefficients of T cells in lymph nodes have been estimated
in several studies employing \vivo two photon microscopy
\cite{Miller.pnas03,Miller.jem04}. Depending on type of cells,
presence of the antigen and activation status of cells, the motility
coefficient has been estimated to range from 10 to 100 $\mu m^2/min$
\cite{Miller.s02,Miller.jem04}. The average size of a mouse lymphocyte
is about $7-10\ \mu m$ \cite{Goldsby.b02}. Assuming that motility of
activated effectors is higher than that of targets, we let $D_E=100\
\mu m^2/min$ and $D_T=10\ \mu m^2/min$. Given that targets used in our
experiments are splenocytes, we also let $R_T=R_S=4\ \mu m$, and for
effectors $R_E=10\ \mu m$. Then the diffusion limited estimate for the
killing efficacy of CD8$^+$T cells \vivo is given by

\beq k = {3(100+10)(10+4)\over 4^3} = 72.2\ min^{-1}.
\label{eqn:kD-final} \ee

However, many of the parameter values are unknown for LCMV-specific
CD8$^+$T cells, and changes in these values can affect the estimate of
the killing efficacy $k$ dramatically. For example, if motility
coefficient for effector T cells $D_E=10\ \mu m/min$, then $k=13.1\
min^{-1}$ which is still several fold higher than the value in
\eref{kD-final}.

\subsection{Alternative ways of fitting the data on killing following
  \vivo infection}

It is generally unknown what killing terms one should use to describe
the process of killing of targets by effector and memory CD8$^+$T cells.
Assuming that killing of targets in a given mouse is determined the
frequency of epitope-specific CD8$^+$T cells in that mouse led to a poor
description of the data (see Main text). Therefore, we investigated
whether using different killing terms can improve the model fit to
data.

Including a decrease in the death rate of targets with an increasing
frequency of targets (see \eref{saturation_T}) failed to improve the
quality of the model fit to data (F-test for nested models,
$F_{1,187}=0.001$, $p=0.98$).  This was not surprising since this
change in the killing term predicts an increased rate of loss of
peptide-pulsed targets with time as more targets are killed, but the
opposite trend is observed in the data (e.g., \fref{fits}C and
\fref{killing_terms}).

We tested whether the average frequency of epitope-specific CD8$^+$T
cells in all mice, rather than values measured in individual mice,
would be predictive of the rate of killing of peptide-pulsed targets.
Therefore, we fitted the data on killing of peptide-pulsed targets
using the mass-action type term $K=kE$ where $E$ is the average
frequency of epitope-specific CD8$^+$T cells in acutely infected or
LCMV-immune mice. This resulted in a significantly better fit with
reasonably small confidence intervals for the estimates of the model
parameters (lack of fit test: $F_{30,162}=0.79$, $p=0.77$; see also
\fref{fits}).

We allowed the death rate of peptide-pulsed targets due to CD8$^+$T cell
mediated killing to saturate with the measured frequency of
epitope-specific CD8$^+$T cells (see \eref{saturation_E}). This also led
to a significantly improved fit of the data (F-test for nested models:
$F_{1,186}=50.3$, $p=2.7\times10^{-11}$). The fit predicted very high
maximal killing efficacy of effector CD8$^+$T cells ($k_{NP396}=146\
min^{-1}$, $k_{GP276}=22\ min^{-1}$, $c_E=400.4$; compare these
estimates to values for effectors given in \tref{parameters}).
Moreover, this model predicted that memory CD8$^+$T cells are only 6\%
(NP396) or 14\% (GP276) as efficient as effectors.  Such an
improvement of the fit by including saturation of the death rate of
peptide-pulsed targets with CTL frequency is expected if measurements
of the frequency of epitope-specific CD8$^+$T cells are noisy.

Finally, by allowing the death rate of peptide-pulsed targets to
depend on the ratio of the frequency of killers to targets (see
\eref{ratio}) we could also obtain an improved fit of the model to
data (lack of fit test: $F_{28,158}=1.38$, $p=0.11$). Interestingly,
we found a relatively small estimate for the parameter $c_T$
($c_T=1.37$). Because effector to target ratios are rather high in
most mice (e.g., $E/\tau \approx 10^3$ for NP396- and $E/\tau \approx
10^2$ for GP276-specific effectors and epitope-expressing targets, see
\tref{parameters}), a small estimate for the constant $c_T$ suggests
that the death rate of peptide-pulsed targets saturates with the
frequency of CD8$^+$T cells (compare \eref{ratio} for $E\gg c_T\freq$ and
\eref{saturation_E} for $c_E E\gg 1$).  Therefore, the last two models
appear to be similar with respect to these \vivo data since both
models require saturation in the death rate of peptide-pulsed targets
with the frequency of epitope-specific CD8$^+$T cells for a satisfactory
description of the data. Such a saturation can simply result from the
reduction of the influence of variation of the measured frequency of
epitope-specific effector and memory CD8$^+$T cells on the death rate of
peptide-pulsed targets.

\setcounter{figure}{0}
\renewcommand{\thefigure}{S\arabic{figure}}

\begin{figure}
\begin{center}
\includegraphics[width=.99\textwidth]{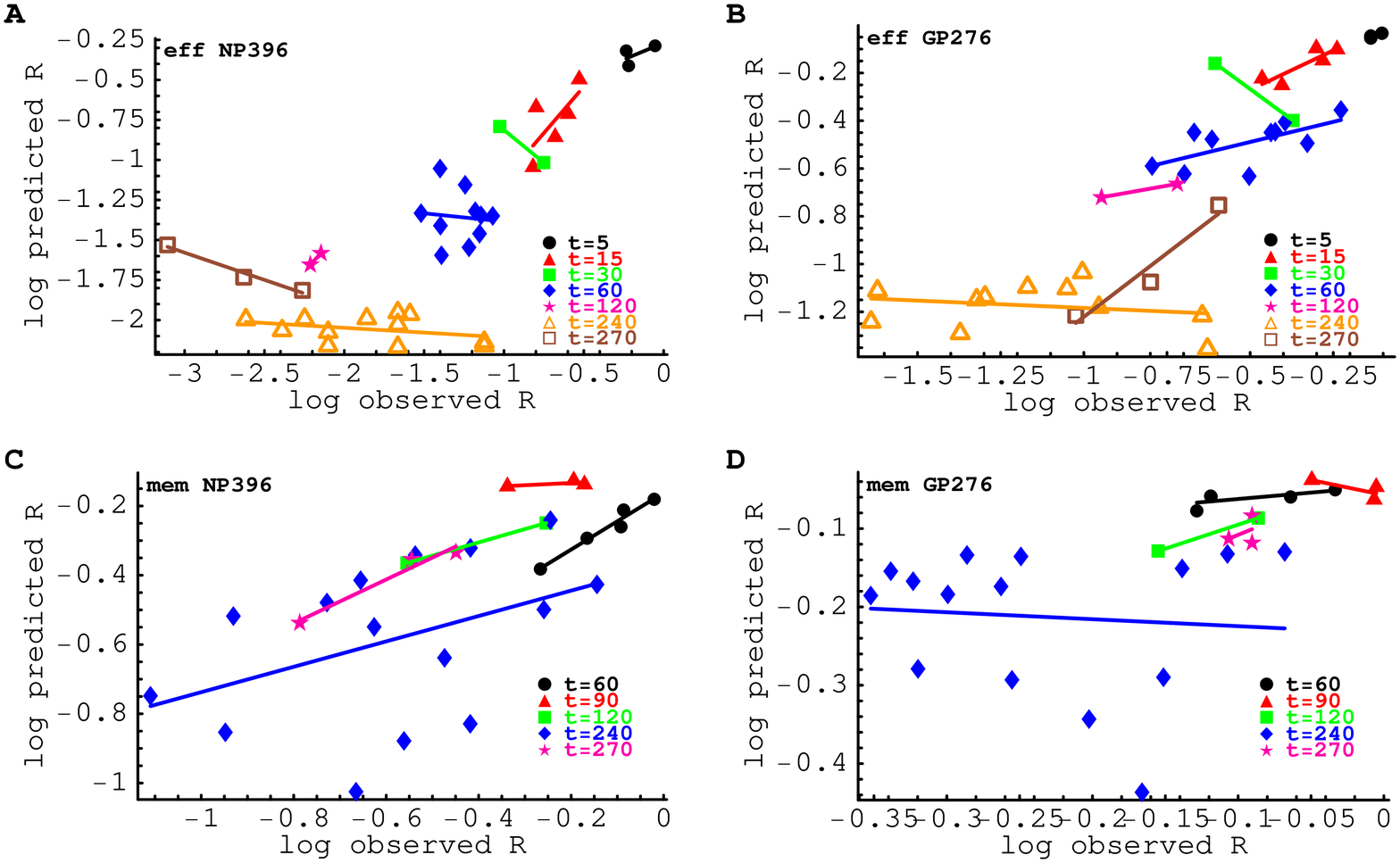}
\end{center}
\caption{Poor prediction of the model taking into account individual
  variation in the measured frequency of epitope-specific effector and
  memory CD8$^+$T cells. We fit the model, that predicts that mice with
  more epitope-specific CD8$^+$T cells should lead to higher killing, to
  the data. Here we plot the log ratio of the frequency of
  peptide-pulsed to unpulsed target cells that is observed in the data
  versus the log ratio that is predicted by the best fit of the model,
  for different times after transfer of target cells. Lines show
  linear regressions.  If the model were to predict the data, the
  points are expected to lie on a line with a positive slope. Instead,
  we observe a large scatter, and for many data, a negative
  correlation between the observation and the prediction.
}\label{fig:prediction-ind-killing}
\end{figure}

\begin{figure}[p]
\begin{center}
  \includegraphics[width=.90\textwidth]{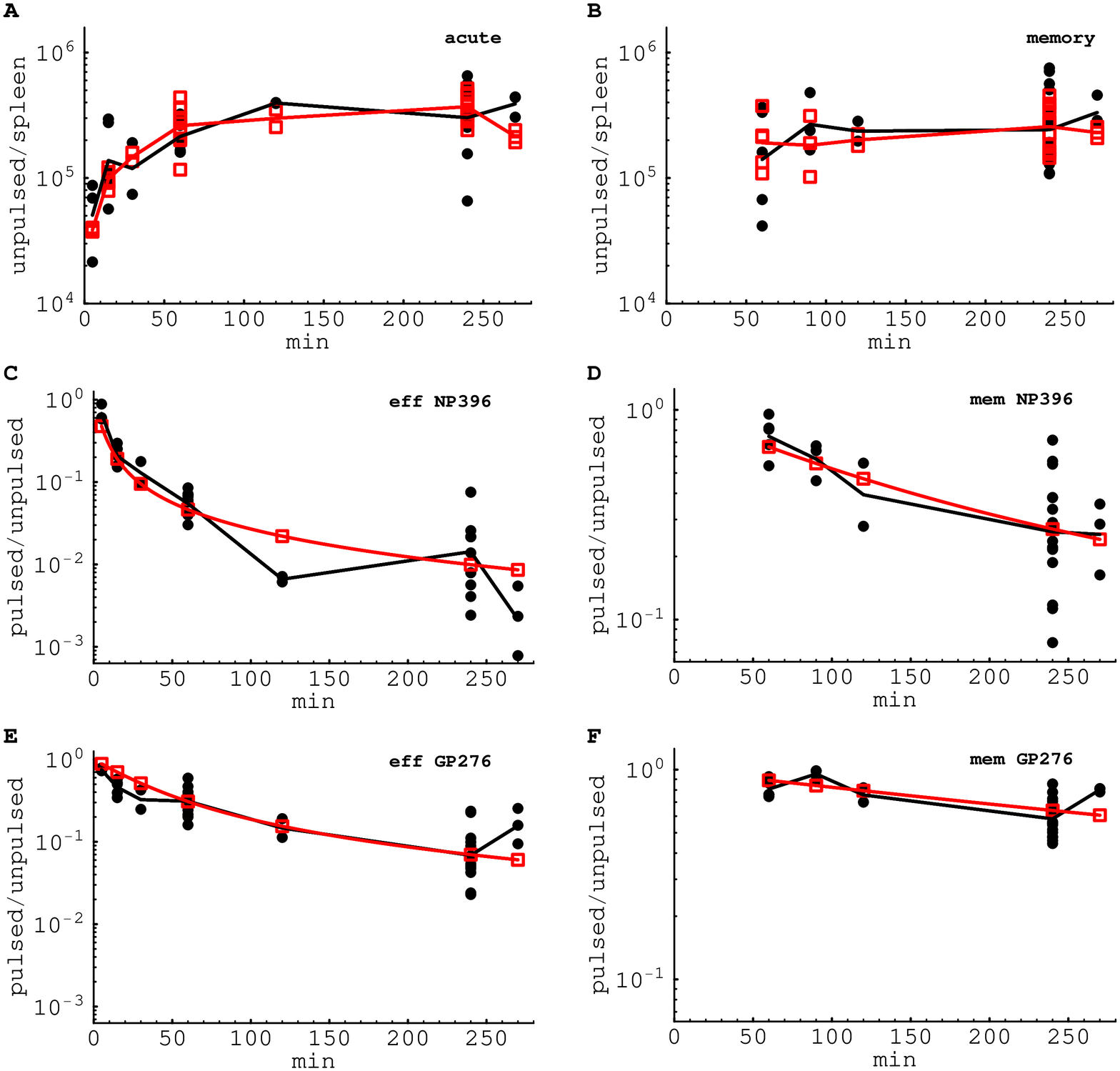}
\end{center}
\caption{Fits of the mathematical model given in
  \eref{S}--(\ref{eqn:R}) to data on killing of NP396- and
  GP276-pulsed targets by effector and memory CD8$^+$T cells.
  The model assumes that the rate of recruitment of targets into the spleen
  depends on the spleen size and that killing of peptide-pulsed
  targets depends on the average frequency of epitope-specific CD8$^+$T
  cells.  Panels A-B show the recruitment of unpulsed targets
  into the spleen, and panels C-F show the decline in the the ratio of
  the frequency of peptide-pulsed to unpulsed targets over time.
  Panels A, C, and E are for acutely infected mice, and panels B, D, F
  are for LCMV-immune (memory) mice.  Panels C and D are for
  NP396-pulsed targets and panels E and F are the GP276-pulsed
  targets. Black dots ($\bullet$) denote measurements from individual
  mice, and black lines denote the log average value per time point.
  Red boxes (\textcolor{red}{$\mathbf \Box$}) show the number of
  recruited cells predicted by the model for individual mice (panels A
  and B) or the predicted average ratio $R$ (panels C-F).  Red lines
  show the log average between individually predicted values.  Note
  the different scale for killing of target cells in acutely infected
  (panels C and E) and memory (panels D and F) mice.  Parameters
  providing the best fit of the model are shown in \tref{parameters}.
  The lack of fit test confirms good quality fits of the data (after
  removing two outliers, $F_{30,162}=0.79$, $p=0.77$). Because of the
  reduced number of parameters, these fits of
  the data are only moderately worse than those obtained in our
  previous study \cite{Ganusov.jv08}.
}\label{fig:fits}
\end{figure}

\begin{figure}
\begin{center}
\includegraphics[width=.99\textwidth]{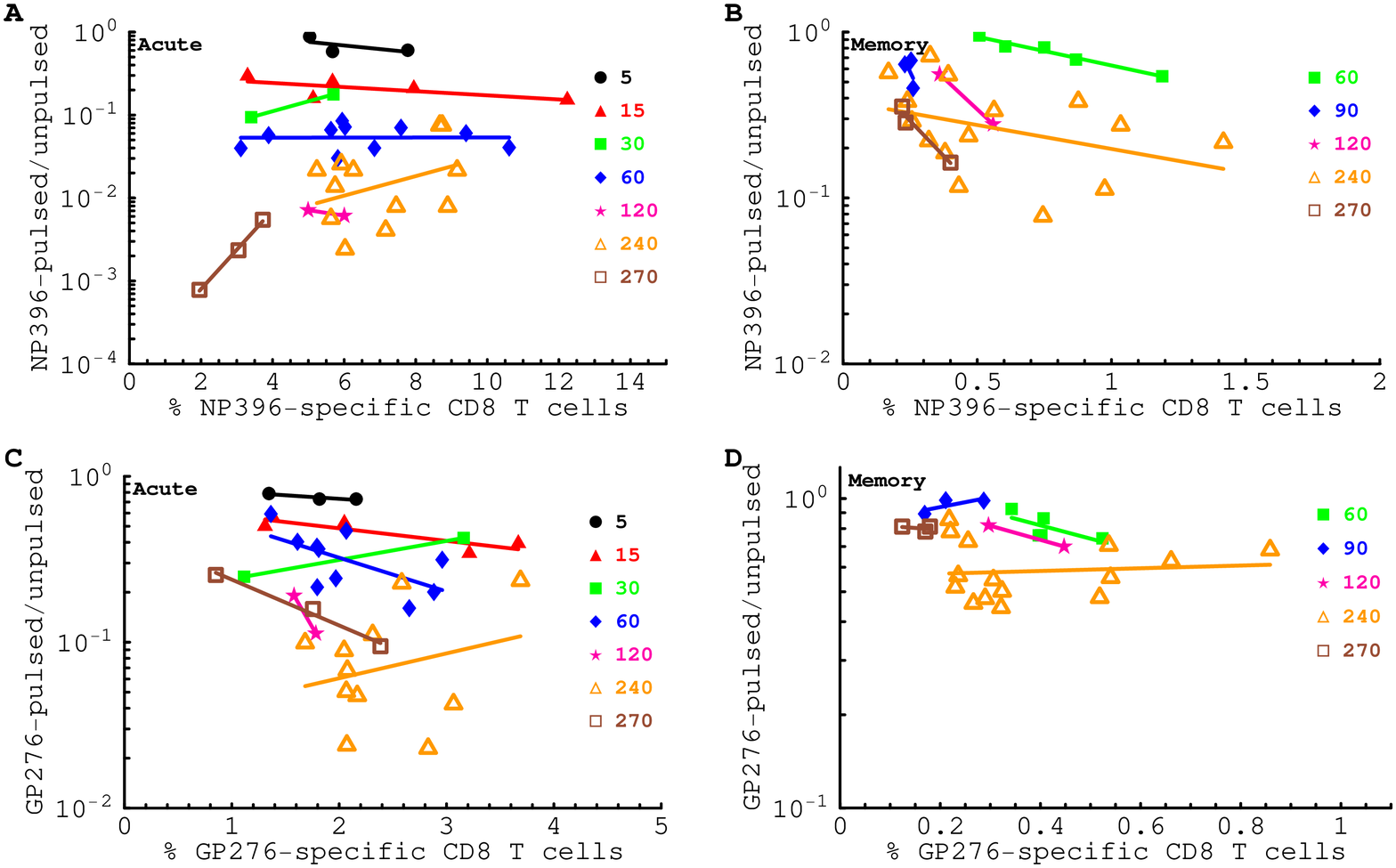}
\end{center}
\caption{The absence of a strong correlation between the ratio of the
  frequency of peptide-pulsed to unpulsed targets at different time
  points after target cell transfer (shown in minutes) and the percent
  of epitope-specific CD8$^+$T cells in the mouse spleen in acutely
  infected (panels A and C) and LCMV-immune (panels B and D) mice. To
  visualize the data, we use different scales on the plots.  If CD8$^+$T
  cells were to affect the frequency of peptide-pulsed targets, a
  negative correlation between the ratio $R$ and CD8$^+$T cells would be
  expected.  However, especially in acutely infected mice (panels A
  and C), there often are positive correlations between this ratio and
  the frequency of peptide-specific CD8$^+$T cells.
}\label{fig:ratio_vs_cd8}
\end{figure}

\begin{figure}
\begin{center}
\includegraphics[width=.5\textwidth]{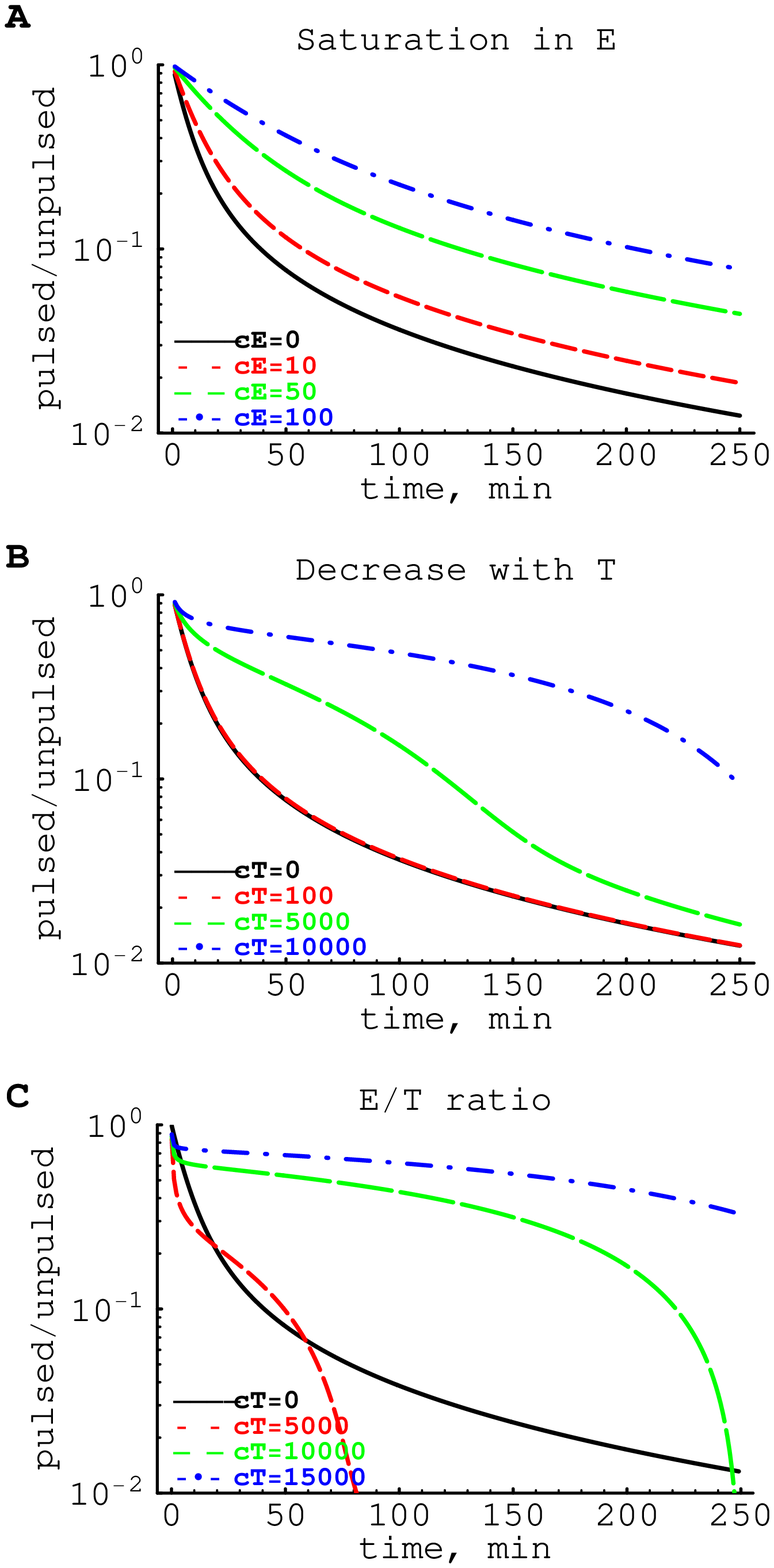}
\end{center}
\caption{Changes in the ratio of the frequency of peptide-pulsed to
  unpulsed targets in the mouse spleen as predicted by the
  mathematical model (given in \eref{S}-(\ref{eqn:R})) with different
  killing terms (given in \eref{saturation_E} -- (\ref{eqn:ratio})).
  The death rate of peptide-pulsed targets saturates with the
  frequency of peptide-specific CD8$^+$T cells (panel A), decreases with
  the frequency of target cells (panel B) or saturates on the effector
  to target ratio (panel C).  We solve the mathematical model
  analytically (panel A) or numerically (panels B-C) with the
  following parameters: $S_B(0)=T_B(0)=5\times10^6$, $S(0)=T(0)=0$,
  $\delta=0.001\ min^{-1}$, $\sigma=0.001\ min^{-1}$, $\eps=0.005\
  min^{-1}$, $k=5\ min^{-1}$, $E=0.05$ (see also \tref{parameters}).
  The frequency of targets in the spleen is calculated as
  $\freq=T(t)/N_s$ where $N_s=8\times10^7$ is the number of
  splenocytes. In all panels, solid lines (with $c_E=c_\freq = 0$)
  predict changes in the ratio $R$ if killing follows the law of
  mass-action, i.e., $K=k E$.  }\label{fig:killing_terms}
\end{figure}

\begin{figure}
\begin{center}
\includegraphics[width=.5\textwidth]{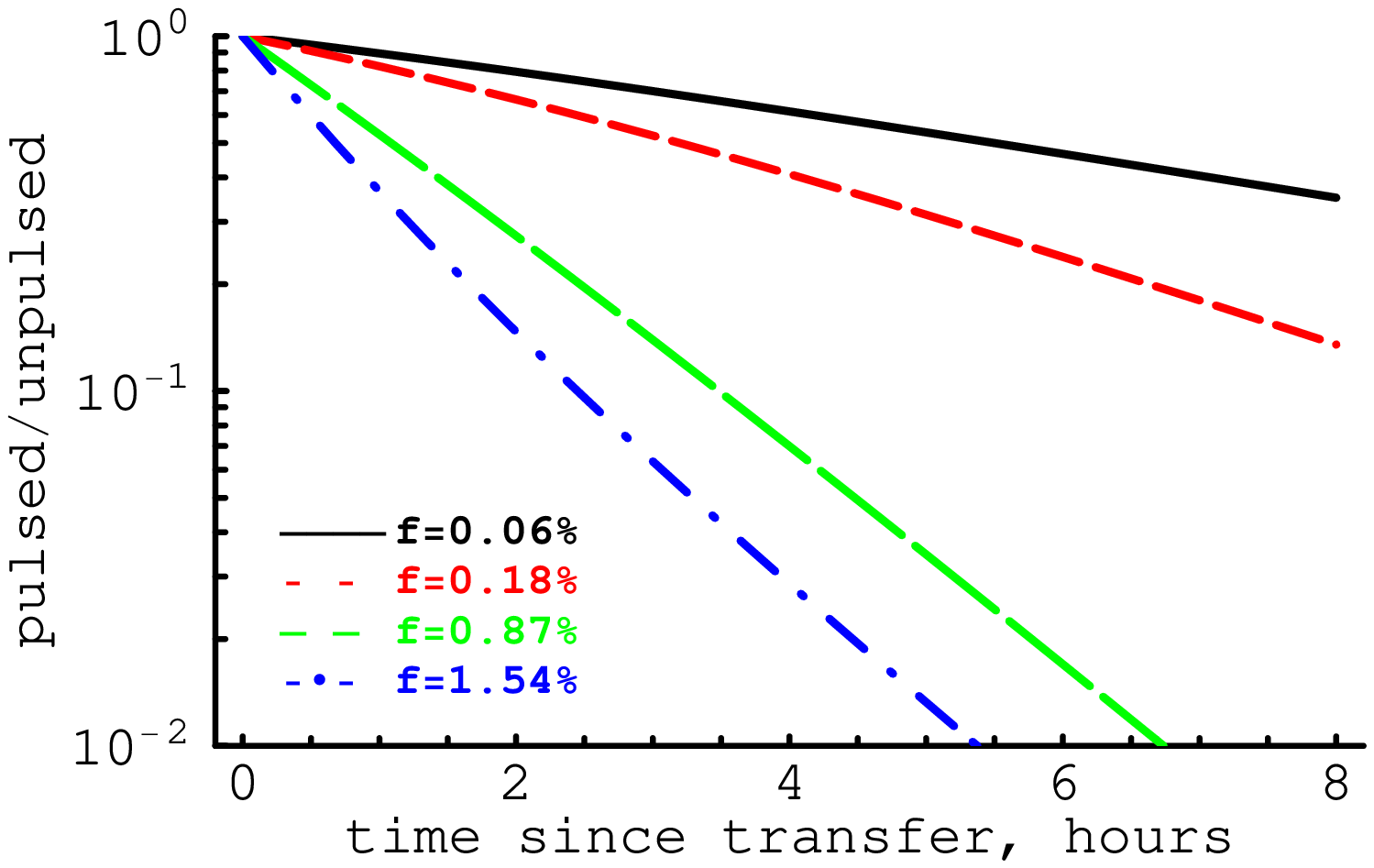}
\end{center}
\caption{Changes in the ratio of the frequency of peptide-pulsed to
  unpulsed targets in the mouse spleen as predicted by the
  mathematical model (given in \eref{R}) for different frequencies of
  GP33-specific effectors in the spleen (shown as $f$).  The death
  rate of peptide-pulsed targets $K$ is proportional to the frequency
  of GP33-specific CD8$^+$T cells, $K=kf$, with $k=2.1\ min^{-1}$ as
  estimated from the adoptive transfer experiments. Other parameters
  are $\al = 2.07\times10^{-11}\ min^{-1}$, $N_s=7.7\times10^8$,
  $\eps=1.15\times10^{-3}\ min^{-1}$, and $\delta=10^{-2}\ min^{-1}$.
  There is a small difference in the percent targets killed between
  $10^6$ and $2\times10^6$ (and between $10^7$ and $2\times10^7$)
  effectors transferred.  }\label{fig:prediction-R}
\end{figure}

\begin{figure}
\begin{center}
\includegraphics[width=.8\textwidth]{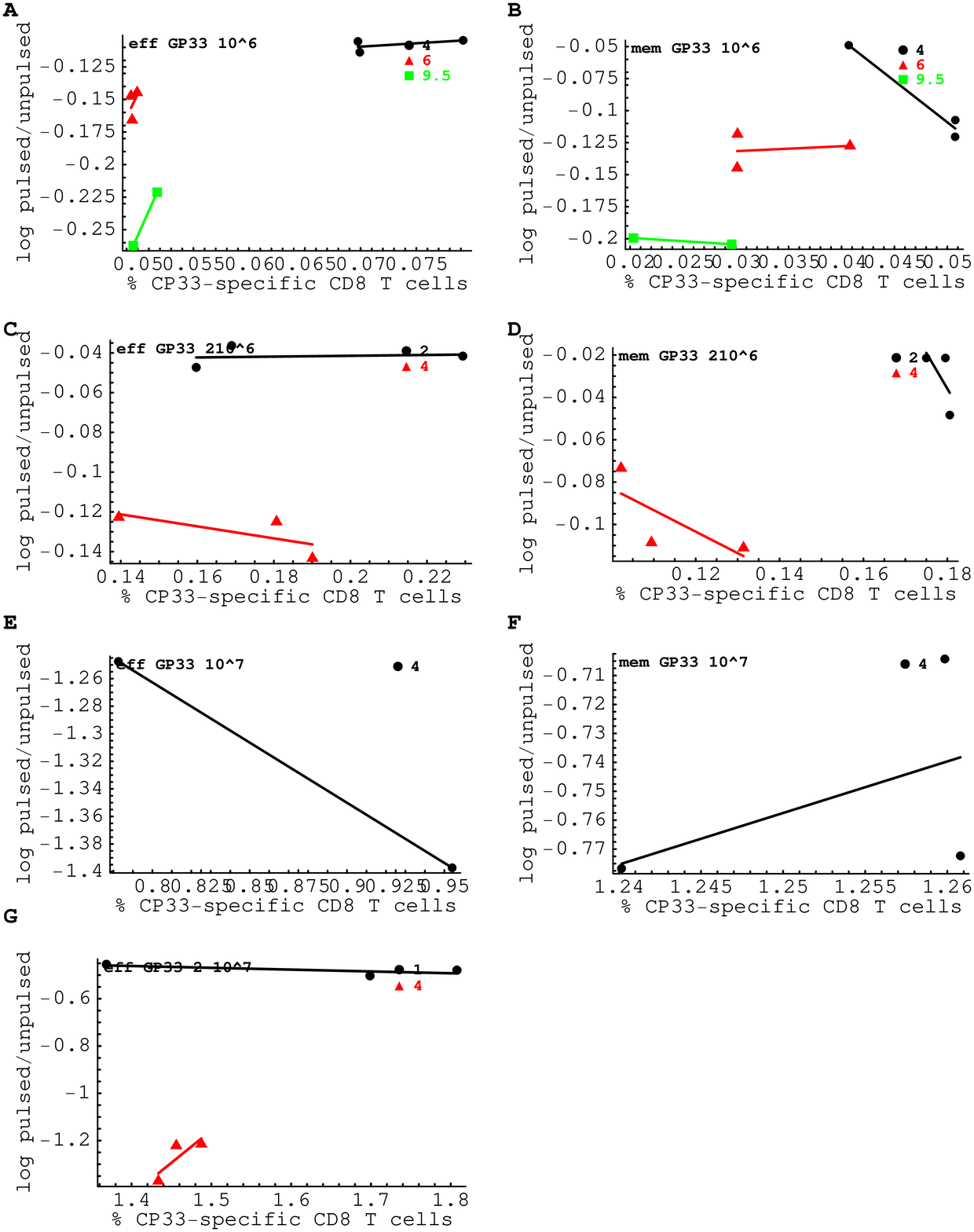}
\end{center}
\caption{The observed correlation between the ratio of the frequency
  of GP33-pulsed and unpulsed targets and the frequency of
  GP33-specific CD8$^+$T cells in the spleen at different time points
  after transfer in the adoptive transfer experiments. Data are
  structured by the time since the transfer of target cells (shown in
  hours). Lines show linear regressions. Note that there
  is a relatively small variation in th measured frequency of
  epitope-specific CD8$^+$T cells.
}\label{fig:R-vs-CD8-transfer}
\end{figure}

\end{document} %